\documentclass[12pt]{article}

\begin{document}

\baselineskip=18.6pt plus 0.2pt minus 0.1pt

 \def\be{\begin{equation}}
  \def\ee{\end{equation}}
  \def\bea{\begin{eqnarray}}
  \def\eea{\end{eqnarray}}
  \def\nn{\nonumber\\ }
\newcommand{\nc}{\newcommand}
\nc{\al}{\alpha}
\nc{\bib}{\bibitem}
\nc{\cp}{\C{\bf P}}
\nc{\la}{\lambda}
\nc{\C}{\mbox{\hspace{1.24mm}\rule{0.2mm}{2.5mm}\hspace{-2.7mm} C}}
\nc{\R}{\mbox{\hspace{.04mm}\rule{0.2mm}{2.8mm}\hspace{-1.5mm} R}}

\begin{titlepage}
\title{
\begin{flushright}
 {\normalsize \small
IFT-UAM/CSIC-04-34 }
 \\[1cm]
 \mbox{}
\end{flushright}
{\bf   On  Geometric Transitions in  String Compactifications
 }\\[.3cm]
\author{Adil Belhaj\thanks{{\tt adil.belhaj@uam.es}}\\[.3cm]
{\it \small  Departamento  de  F\'{\i}sica Te\'orica, C-XI, 
Universidad Aut\'onoma de Madrid } \\ {\it \small Cantoblanco,
E-28049-Madrid, Spain}
\\
{\it \small  
Instituto de F\'{\i}sica Te\'orica, C-XVI,
Universidad Aut\'onoma de Madrid } \\ {\it \small Cantoblanco,
E-28049-Madrid, Spain}
\\[.2cm]
  } } \maketitle
\thispagestyle{empty}
\begin{abstract}

We  reconsider the  study   of the  geometric transitions and
brane/flux dualities  in  various dimensions.   We first  
  give  toric interpretations of the topology changing transitions in the 
  Calabi-Yau  conifold and the $Spin(7)$ manifold.  The latter, for instance,
can be viewed as three intersecting Calabi-Yau conifolds according  to  $\cp^2$ toric graph. 
 Orbifolds of such geometries are given in terms of  del Pezzo complex surfaces. Second we  
 propose a four-dimensional  F-theory interpretation of type IIB
 geometric transitions  on the  Calabi-Yau conifold. This   gives  a dual  description of the
M-theory flop  in terms of  toric mirror symmetry.   In two dimensions,  we study
  the  geometric  transition  in a  singular $Spin(7)$ manifold
constructed as a cone on SU(3)/U(1). In particular, we discuss  
brane/flux duality  in such a compactification  in both type IIA and type IIB superstrings. These examples preserve one 
supercharge and so 
 have   ${\cal N}= 1/2$ supersymmetry in two dimensions. Then,  an interpretation in terms of F-theory is given.
\end{abstract}
{\tt  KEYWORDS}: Toric geometry, Mirror symmetry,   String theory,
   Manifolds with non trivial holonomy groups, and Geometric transitions.

\end{titlepage}

\tableofcontents

\newpage

\section{Introduction}

It has been known for a long time that duality plays an
interesting role in the context of string theory. More recently,
geometric transitions have become a tool in understanding large
$N$ dualities between $SU(N)$  gauge theory  and
 closed string models \cite{GV,V, BR}. Some well
known examples are provided by D-branes wrapped on cycles in
manifolds with non trivial  holonomy groups, where a good description is given by the
small coupling limit of the corresponding worldvolume gauge
theory.  After the geometric transition, the D-branes disappear
and they are replaced by fluxes through cycles in the dual
geometry, providing the adequate description of the same physics
at strong coupling~\cite{GV, V}. In  Calabi-Yau geometry, for instance, the large $N$ duality
between D6-branes wrapped around the ${\bf S}^3$ of the deformed
conifold, $T^\ast {\bf S}^3$, and type IIA superstring on the small
resolution of the conifold, $O(-1)+O(-1)$ bundle over $\cp^1$,  with fluxes has been studied
in~\cite{V}. This result has been 'lifted' to 
M-theory \cite{AMV} where it corresponds to the so-called
flop duality in M-theory compactified on a manifold
with $G_2$ holonomy, for short, a $G_2$ manifold.  The mirror version of  type
 IIA duality becomes the large
$N$ equivalence between D5-branes wrapped on the ${\cp}^1$ of
the resolved conifold and type IIB superstring  with three-form fluxes through
the ${\bf S}^3$ of $T^\ast {\bf S}^3$.  This has been studied and extended to
more general Calabi-Yau geometries  \cite{CIV, Ta, CKV, CFIKV, AgMV, G, LL,AABDS}.
\par
Recently, similar studies have been done in three and  two 
dimensions using  respectively  type IIA superstring compactified 
on a  $G_2$ manifold
  \cite{AW,GST},
and  type IIA( or B) on an eight-dimensional manifold with
   $Spin(7)$  holonomy group, for short,  $Spin(7)$ manifold
\cite{GST, BR}. This $Spin(7)$
manifold is constructed as a cone on $SU(3)/U(1)$ \cite{J} and has
a  geometric transition involving a collapsing
${\bf S}^5$ and a growing $\cp^2$. In type II superstrings, this could   be interpreted
as a transition between two phases described by wrapped
D-branes or R-R fluxes \cite{BR}.

The aim of this work is  to  reconsider the  study   of the  geometric transitions and
D-brane/flux dualities  in  various dimensions.   We first  
  give  toric interpretations of the topology changing transitions in  the 
  Calabi-Yau  conifold and the $Spin(7)$ manifold.  The latter, for instance,
can be viewed as three intersecting Calabi-Yau conifolds according  to  $\cp^2$ toric graph. 
 Orbifolds of such geometries are given in terms of  del Pezzo complex surfaces. Second we  
 propose a four-dimensional  F-theory interpretation of type IIB
 geometric transitions  on the  Calabi-Yau conifold. This   gives   a dual description of the
M-theory flop  in terms of  toric mirror symmetry.   In two dimensions,   we study
  the  geometric  transition  in a  singular $Spin(7)$ manifold
constructed as a cone on SU(3)/U(1). In particular, we discuss  
brane/flux duality  in such a compactification  in both type IIA and type IIB superstrings.  
Then, we engineer  gauge symmetries going beyond the models given in \cite{BR}.
These examples preserve one 
supercharge and so 
 have   ${\cal N}= 1/2$ supersymmetry in two dimensions.

The  organization of this  work is  as follows. In section 2, we
briefly review the main line  of toric geometry method for
treating  manifolds with tori fibrations.  In section 3, we 
 give a toric interpretation of the topology changing transitions in 
such manifolds. In particular,  we  discuss the Calabi-Yau
conifold transition,  the $Spin(7)$
manifold and  its   generalization to  non-trivial 
geometries.  In section 4,   we  reconsider the study  of the transition in 
four dimensions  by giving a new F-theory interpretation  of the transition duality
in  type IIB  superstring theory  compactified on the Calabi-Yau conifold.
In the case  of  type II
superstrings on the  $Spin(7)$ manifolds, we  give a conjecture on
 D-brane/flux duality
in  type IIA and type IIB  with  ${\cal N}=1/2$ in two dimensions.  This is given in  section 5.
We end this study with discussion  and  some open questions.

\section{Toric geometry}
\subsection{Projective spaces and odd-dimensional spheres}
 In  this section,  we  collect a  few facts  on  toric  realizations of 
 non-trivial geometries. These facts   are  needed  later  on when
 discussing the topology changing  in 
 manifolds with non-trivial holonomy groups.
   Roughly speaking, toric manifolds are, in general,  complex
  $n$-dimensional manifolds
   with $T^n$ fibration over real  $n$-dimensional  base spaces with boundary
  \cite{LV, BS1, AB}.
    They exhibit
    toric actions $U(1)^n$  allowing us   to encode the
  geometric properties of the complex spaces in terms of simple
  combinatorial data of polytopes ${\Delta}_n$ of the $R^n$ space.
  In this correspondence, fixed points of the toric actions $U(1)^n$
  are associated to the vertices of the polytope ${\Delta}_n$, the
  edges are fixed one-dimensional lines of a subgroup $U(1)^{n-1}$
  of the toric action $U(1)^n$, and so on.   \par   To illustrate the main idea of  toric geometry, let us
  describe  the philosophy of this subject  through certain useful  examples, and then we give some generalizations useful later on.\\
   { (i) $\cp^1$  projective space}\\
   The simplest example, in toric geometry, is probably  $\cp^1$. This manifold  
   plays a crucial role in  the  building
   blocks  of higher-dimensional toric
 varieties    and  in the study of the small resolution of 
singularities of  local  Calabi-Yau manifolds. Roughly, 
 $\cp ^1$ has an
  $U(1)$ toric action
 \be  z\to e^{i \theta}z\ee with  two fixed points
  $v_1$ and $v_2$ on the real line. The latter  points,  which  can
  be generally chosen as $v_1=-1 $ and $v_2=1$, describe
  respectively north and south poles of the real two sphere $ S^2
  \sim \cp^1$. The corresponding one-dimensional polytope is just
  a  segment $[v_1,v_2]$ joining the two points $v_1$ and $v_2$.
  Thus, $\cp^1$ can be viewed as  the  segment $[ v_1,v_2]$  with a
  circle on top,  where the circle vanishes at the end points $v_1$
  and $v_2$.
\be
      \mbox{
         \begin{picture}(20,20)(40,0)
        \unitlength=2cm
        \thicklines
    \put(0,0){\line(1,0){2}}
     \put(-0.25,0){$v_1$}
     \put(2.1,0){$v_2$}
  \end{picture}
}
\label{line}
\ee
 { (ii) $\cp^2$ projective space}\\
       This  is a complex  two-dimensional toric variety defined by
  \begin {equation}
  {\cp^2}= {C^{3}\setminus\{(0,0,0)\}\over C^*}, 
  \end{equation}
  where  ${C^*}:\;\; (z_1,z_2,  z_{3}) \to ( \lambda z_1,\lambda z_2,\lambda z_{3})$.
  This manifold  admits   an  $U(1)^2$ toric  symmetry
 acting 
as follows
\bea  (z_1,z_2,  z_{3}) \to ( e^{i
  \theta_1} z_1,e^{i \theta_2} z_2, z_{3})
\eea  and exhibiting three
  fixed points $v_1$, $v_2$ and $v_3$. The corresponding polytope
  ${\Delta}_2$ is a finite sublattice of the $ \bf Z^2$ square lattice.
  It describes the intersection of three $\cp^1$'s   defining a
  triangle ($v_1 v_2 v_3$) in the $ \bf R^2$ plane.  Thus,  ${\Delta}_2$ has three edges,
  namely $[v_1,v_2]$, $[v_2,v_3]$ and $[v_3,v_1]$, stable under the
  three $U(1)$ subgroups of $U(1)^2$. Two subgroups are just the two
  $U(1)$ factors,  while the third subgroup is the diagonal one.
 $\cp^2$ can be viewed as a triangle over
  each point of which there is an elliptic curve  $ T^2$. This torus
 shrinks to a
  circle at each segment   $[v_i,v_j]$  and  it shrinks to a  point at
 each $ v_i$. This is shown in the following figure
\be
\mbox{
 \begin{picture}(100,150)(0,-15)
    \unitlength=2cm
  \thicklines
   \put(0,0){\line(1,2){1}}
   \put(0,0){\line(1,0){2}}
  \put(2,0){\line(-1,2){1}}
 \put(0.9,2.1){$ v_1$}
 \put(-0.25,0){$v_2$}
   \put(2.1,0){$v_3$}
   \put(-0.1,1.1){$z_3=0$}
  \put(0.7,-0.3){$z_1=0$}
   \put(1.55,1.1){$z_2=0$}
\end{picture}
}
\label{triangle}
\ee
 This  toric realization will play a crucial role in our present  work.  For $n$-dimensional
projective spaces $\cp^n$, the  toric representation   is  given by  ${\bf T}^n$ fibration over
a  $n$-dimensional simplex.\par 
 The above representation  can be extended to   some real manifolds, 
in particular the odd-dimensional real spheres   being related to  $\cp^n$ by
\be
\cp^n={\bf S}^{2n+1}/{\bf S}^1.
\ee
 In this way, ${\bf S}^{2n+1}$
is a ${\bf S}^1$ bundle over $\cp^n$.  Using this   realization,  one can
 give a toric representation for
 odd-dimensional spheres. Indeed,
the one-sphere, for example, is trivially realized as a ${\bf
T}^1\sim{\bf S}^1$ over the zero-simplex -- a point. As we have
seen, the three-sphere may be realized as a ${\bf T}^2$ over a
one-simplex -- a line segment as the one in (\ref{line}). This may
be extended to the $(2n+1)$-dimensional sphere ${\bf S}^{2n+1}$
which may be described as a ${\bf T}^{n+1}$ over an $n$-simplex.
Of particular interest in this work  is the five-sphere ${\bf S}^5$ being  realized as the triangle (\ref{triangle}) with a
${\bf T}^3$ on top (whereas $\cp^2$ had a ${\bf T}^2$ on top).

\subsection{  More  general toric   varieties}
 Let us now consider a more  complicated example. The geometries 
 that we will study  in this subsection 
  can be also  described as  quotient spaces.    Consider the complex space  ${C^{n+r}}$ 
parametrized  by $z_1,...,  z_{n+r}$ and $r$ ${C^*}$ actions  given by
  \begin {equation}
\label{toric}
   {C^*}^r: z_i   \to \lambda^{Q_i^\alpha} z_i,\;\; i=1,2,\ldots, n+r,  \quad \alpha=1,
  2,\ldots,r,
  \end{equation}  where  $Q_{i}^{\alpha}$ are integers. For each $\alpha$ they form
the so-called  Mori
  vectors in toric  geometry. They   generalize  the  weight vector
$(w_i)$ of the  complex  $n$-dimensional  weighted projective space
 ${\bf W}\cp^n_{w_1,\ldots,w_{n+1}}$. Now
one can define  a general toric variety  ${{\cal \bf V}^n}$  by the following  symplectic
quotient space 
  \begin {equation}
  {{\cal \bf V}^n} = {C^{n+r}\setminus U\over {C^*}^r},
  \end{equation}
     where  $U$ is a subset  of
  $C^{k}$ chosen by triangulation \cite{BS1}.
  \\
   ${{\cal \bf V}^n}$  has a   $ T^n$  fibration,
 obtained
    by dividing   $ T^{n+r}$  by the $U(1)^r$  gauge symmetry
    \be
    z_i\to e^{iQ^\alpha_i \vartheta^\alpha} z_i,\quad \alpha=1,\ldots,r,
    \ee
   where $\vartheta^\alpha$ are the generators of the $ U(1)$ factors.
It   can
 be  represented by a  toric
     graph  $ \Delta({{\cal \bf V}^n})$ spanned by  $ k=n+r$
     vertices $ v_i$ in  $\bf Z^n$ lattice satisfying
      \begin {equation}
    \sum \limits _{i=1}^{n+r} Q_i^\alpha v_ i=0,\quad \alpha=1,\ldots,r.
  \end{equation}
 This  geometric description   of ${{\cal \bf V}^n}$  has 
 a    nice physical realization through  the ${\cal N}=2$ linear sigma model.
 The theory has  an
$U(1)^r$ gauge  symmetry with   $n+r$ chiral fields $\phi_i$  and a   $Q_i^\alpha$
matrix gauge charge  ~\cite{W}. In this way, ${{\cal \bf V}^n}$ is a solution 
of the vanishing condition of  the D-term potential ($D^\alpha=0$), up to $U(1)^r$ gauge
transformations, namely
\begin {equation}
  \label{MinimumDTerm}
  \sum \limits _{i=1}^{n+r} Q_i^\alpha|\phi_i|^2=\rho_\alpha,
\end{equation}
where the $\rho_\alpha$'s are Fayet-Iliopoulos (FI) coupling parameters.
The (local) Calabi-Yau condition is satisfied by
\begin {equation}
  \sum \limits _{i=1}^{n+r} Q_i^\alpha=0, \quad \forall \alpha,
\end{equation}
which means that the physical system flows in the IR to a
non-trivial superconformal theory~\cite{W,BS2}. \par
Finally note  that for a given toric complex manifold,  one  can
 construct its mirror  using
   two-dimensional  sigma model analysis \cite{HV,A}.   Indeed,
 the mirror version  of the constraint equation (\ref{MinimumDTerm}),    giving the
 superpotential in the Landau-Ginsburg (LG) models,  reads
\be
\sum\limits_i  a_iy_i=0
\ee
subject to
\be
\prod \limits_i y_i^{Q_i^\alpha}=1. \ee
 In these equations, $ y_i$ are  LG  dual  chiral fields which can be related, up to some
field changes, to sigma model fields and where  $a_i$'s 
   can  be identified  with the
complexified FI parameters,  defining  now the complex deformations
of  the LG Calabi-Yau superpotentials.

\section{Toric  varieties    and geometric transitions}
 In this  section, we  reconsider the study  of the topology changing 
   of manifolds with exceptional
 holonomy group using toric geometry. In particular, we  discuss
 the case of the  Calabi-Yau manifold  and
$Spin(7)$ manifolds.

\subsection{Toric Calabi-Yau conifold}
 Let  us   consider  first  the known example corresponding to
 three complex dimensions which is  the so-called  conifold. This manifold
is defined by the following algebraic  equation
\begin{equation}
  \label{ConifoldEqn}
  uv-xy=0
\end{equation}
where the singularity is located at $(u,v,x,y)=(0,0,0,0)$. There are
basically two ways of smoothing this singularity, either by
``toric'' small resolution or by complex deformation. \\

\noindent {\bf\underline{Toric small resolution}:}
It consists in replacing the singular point $(0,0,0,0)$ by a toric $\cp^1$
manifold.  This resolution can be described by a  ${\cal N}=2$  toric sigma model realization. The theory is 
an   $U(1)$ gauge  model  with four
chiral matter fields $\phi_i$ with  the vector charge $Q_i=(1,1,-1,-1)$. 
In this way, the conifold equation~(\ref{ConifoldEqn}) can be
solved in terms of the gauge invariant  terms as follows
\begin{equation}
  u=\phi_1\phi_3\quad v=\phi_2\phi_4 \quad x=\phi_1\phi_4
 \quad y=\phi_2 \phi_3\, .
\end{equation}
Turning on the FI D-terms, which are given by,
\begin {equation}
  V_D= ( |\phi_1|^2+|\phi_2|^2-|\phi_3|^2-|\phi_4|^2-\rho)^2\, ,
\end{equation}
corresponds to blowing up the origin by a  toric  $\cp^1$. Indeed, 
consider, for example, the case $\rho>0$. Then the fields $\phi_1$
and $\phi_2$ cannot be zero, and these two coordinates define a $
\cp^1$ given (up to a $U(1)$ gauge transformation) by
\begin{equation}
  |\phi_1|^2+|\phi_2|^2=\rho\, .
\end{equation}
The fields $\phi_3$ and $\phi_4$ can be regarded as non-compact 
coordinates parameterizing the normal
directions for the fibers. Then the  total space  of the small resolution
is  a toric variety given by  the bundle  $O(-1)+O(-1) \to \cp^1$,
which is topologically $\R^4 \times \bf S^2$.  A similar analysis can
be done for $\rho<0$ by exchanging the role of the base and the
fiber. These two small resolutions ($\rho<0$ and $\rho>0$) of the
conifold are related by the so-called flop transition.\\

\noindent {\bf\underline{Complex deformation}:}
Besides the toric resolution we have discussed above, the conifold
singularity can be deformed by keeping the Kahler structure and
modifying the defining algebraic equation as follows
\begin{equation}
  \label{DeformedConifoldEqn}
  uv-xy=\mu\, ,
\end{equation}
where $\mu $  is a complex  parameter.  Now the singular point is replaced by an $\bf S^3$
 being  obtained by taking a real parameter $\mu$, $u=\bar v$ and
$x=-\bar y$.  Up to changes of variables, one can show that  the   total    geometry
is nothing but  $T^\ast \bf S^3$, whose topology is clearly $  \R^3\times
S^3$.   This is called the deformed conifold and is related to the
resolved conifold by the so-called conifold transition.

The  conifold transition admits a representation
in toric geometry, where it can be understood as an
enhancement or breaking, respectively, of the toric
circle actions. On the one hand, the $O(-1)+O(-1)$ bundle
over $\cp^1$ has only one toric $U(1)$ action, identified
with the toric action on $\cp^1$ itself, while the deformed
conifold $T^\ast{\bf S}^3$ has a toric $U(1)^2$
action since the spherical part can be viewed as a
${\bf T}^2$ over a line segment.
The torus is generated
by the two $U(1)$ actions
\be
 (u,v)\ \rightarrow\ (e^{i\theta_1}u,e^{-i\theta_1}v)\ ,\ \ \ \ \ \ \ \
 (x,y)\ \rightarrow\ (e^{i\theta_2}x,e^{-i\theta_2}y)
\ee
with $\theta_i$ real.
Thus, the blown-up ${\bf S}^3$ may be described by the
complex interval $[0,\mu]$ with the two circles parameterized
by $\theta_i$ on top, where ${\bf S}^1(\theta_1)$
collapses to a point at $\mu$ while ${\bf S}^1(\theta_2)$
collapses to a point at $0$.
The transition occurs when one of these
circles refrains from collapsing while the other one collapses
at both interval endpoints. This breaks the toric
$U(1)^2$ action to $U(1)$, and the missing $U(1)$ symmetry
has become a real line (over $\cp^1$). The resulting geometry
is thus the resolved conifold.

\subsection{Toric   representation   of  the transition in  the  $Spin(7)$ manifold}

In this subsection, we   propose   a picture for understanding the
topology changing  transition of the $Spin(7)$
manifold discussed in \cite{GST} using toric geometry. 
The example we shall be interested in may be described
as a singular real cone over the seven-dimensional
Aloff-Wallach (coset) space $SU(3)/U(1)$.
It was argued in \cite{GST} that there are two ways of
blowing up the singularity, replacing the singularity
by either $\cp^2$ or ${\bf S}^5$. The resulting
smooth $Spin(7)$ manifold is given by
$ \R^4\ \times\ \cp^2$ and  $\R^3\ \times{\bf S}^5$
and are referred to as resolution and deformation, respectively,
due to the similarity with the Calabi-Yau conifold ($ \R^4\ \times\ \cp^1$ and  $\R^3\ \times{\bf S}^3$).

Here we shall  re-consider  the transition between these two
manifolds in the framework of  toric geometry. The basic idea \cite{BR} is to view
the singular $Spin(7)$ manifold (the real cone on $SU(3)/U(1)$)
as three intersecting Calabi-Yau conifolds associated to the
triangular toric
diagram (\ref{triangle}). The deformed and resolved $Spin(7)$
manifolds then correspond to the three intersecting
 Calabi-Yau conifolds being deformed or resolved, respectively.

Indeed, let us consider  the complex linear space  $\C^3$ 
described by the three coordinates
$(z_1,z_2,z_3)$.  Let us introduce the
constraint equation 
\be
 |z_1|^2+|z_2|^2+|z_3|^2\ =\ r
\label{r}
\ee
 where  $r$ is real and positive. This defines a ${\bf S}^5$ while the additional identification
\be
 (z_1,z_2,z_3)\ \sim\ (e^{i\theta}z_1,e^{i\theta}z_2,e^{i\theta}z_3)
\label{theta}
\ee
(with $\theta$ real)
will turn it into a $\cp^2$. In either case, $r$ measures the size.
With both conditions imposed, we can obtain the
three resolved Calabi-Yau conifolds
$
 \R^4\ \times\ \cp^1(z_k=0)\ , k=1,2,3$
  embedded in $\R^4\times\C^3$, simply by setting one of the
coordinates equal to 0. The resolution of the $Spin(7)$
singularity reached by blowing up a $\cp^2$ can thus be described
by three intersecting resolved conifolds over the triangle
(\ref{triangle}). Likewise, the deformation of the $Spin(7)$
singularity constructed by blowing up a ${\bf S}^5$ may be
realized as three intersecting deformed Calabi-Yau conifolds
$\R^3\ \times\ {\bf S}^3(z_k=0),  k=1,2,3$ over the same triangle.
 Note that this  interpretation of the $Spin(7)$ manifold as
three intersecting Calabi-Yau manifolds over a triangle we have given 
corresponds to all three
Calabi-Yau manifolds undergoing simultaneous
conifold transitions.

\subsection{More general geometric transitions}
 There are several generalizations of the 
above  analysis of the
$Spin(7)$ manifold.  We discuss  some of them  briefly  here and leave others  for future research. 

\subsubsection{Orbifolds of $Spin(7)$ manifolds  and del Pezzo  surfaces}
A simple generalization of the above idea is to consider   orbifolds of $Spin(7)$ manifolds. This 
involves   del Pezzo complex surfaces  $dB_k, k=1,2,...$ as base geometries.  First we recall  that  
$dB_k$ are two-dimensional complex surfaces   that are  obtained  by blowing up
to $k$ points  in $\cp^2$.     Alternatively $dB_k$ can be obtained as  Hirzebruch surfaces $F_k$ being spheres
 fibered over spheres. In this way,  $dP_k$ can be identified with $F_{k-1}$. 
With the number
of points restricted as $k=1,2,3$, this defines  so-called the
toric del Pezzo surfaces. In toric geometry, 
the blowing up consists in replacing a point by $\cp^1$ with
a line segment as its toric diagram. The full del Pezzo surface
will thus have a polygon with $k+3$ legs as its toric diagram.  Let us now consider 
the resolved $Spin(7)$ manifold and introduce a $Z_2$  discrete group  acting only on two homogeneous 
 coordinates of  $\cp^2$ as follows
\bea  
(z_1,z_2,z_3) \to ( - z_1,- z_2, z_3).
 \eea 
  This  transformation leads to  a singular geometry.  Locally, near such a singularity, this looks like 
an $A_1$ ALE space. This is  given by 
\be
uv=z^2
\ee
where $u=z_1^2$, $v=z_2^2$ and $z=z_1z_2$. The blow up of this singularity leads to 
$\cp^1\times \cp^1$   which can be identified  with $dP_1$.  So, the total geometry can be 
regarded now as four intersecting resolved conifold according to a rectangle.  After transition, we expect that 
the  same intersection but  with   the deformed  Calabi-Yau conifolds. A similar analysis could be done
for a generic valued  of $k$.  In this way, the resulting geometry is given by $k+3$ intersecting  Calabi-Yau
 manifolds according to a polygon.

\subsubsection{Higher-dimensional geometries}

 Another possible generalization is to consider   $n$-dimensional  projective space $\cp^n$.
Since ${\bf S}^{2n+1}$ can be described as a ${\bf T}^{n+1}$
over an $n$-simplex it supports a toric $U(1)^{n+1}$ action
whereas $\cp^n$ (which may be realized as a ${\bf T}^n$
over an $n$-simplex) admits a toric $U(1)^n$  symmetry. Like Calabi-Yau conifold, we are thus
expecting that a geometric transition can take place,
replacing a $U(1)$ by the one-dimensional real line $\R$.
Since the $U(1)$ is associated to one of the ${\bf S}^1$
factors of ${\bf T}^{n+1}$, the transition essentially amounts to
replacing ${\bf T}^{n+1}$ by ${\bf T}^n\times\R$.
Our interest is in real fibrations over the
spaces ${\bf S}^{2n+1}$ and $\cp^n$
so the relevant geometric transitions  \cite{BR} may be read as 
\be
 (\mbox{deformed geometry})\ \ \ \ \ \ \R^3\ \times\ {\bf S}^{2n+1}\ \ \ \ \longleftrightarrow\ \ \ \
 \R^{4}\ \times\ \cp^n\ \ \ \ \ \ (\mbox{resolved geometry})\ .
\label{m}
\ee 
Relating it to  toric geometry, this transition has a 
representation as intersecting conifolds over an $n$-simplex, provided by a 
 simple combinatorial analysis.  Indeed, the number of intersecting Calabi-Yau
 conifolds in the toric picture is equal to the number of one-dimensional edges
of the simplex, namely $\frac{1}{2}n(n+1)$. Similarly,  one should
also expect to be able to describe the transition in terms of
intersecting $Spin(7)$ manifolds over the
$n$-simplex. In this  case, the number of intersecting
$Spin(7)$ conifolds is equal to the number of two-dimensional
faces of the simplex, $\frac{1}{6}n(n^2-1)$.

\section{Geometric transitions in  four dimensions}

\subsection{Duality  in  type IIA }

Gopakumar and Vafa have recently  conjectured  that the $SU(N)$ Chern-Simons
theory on ${\bf S}^3$ for large $N$
is dual to topological strings on the resolved conifold \cite{GV}.
In this way, the 't Hooft expansion of the Chern-Simons free energy
has been shown to be in agreement, for all genera, with the
topological string amplitudes on the resolved conifold.
This duality has subsequently been embedded in type IIA
superstring theory \cite{V}, where it was proposed that $N$
D6-branes wrapped around ${\bf S}^3$ of the deformed
conifold is equivalent (for large $N$) to type IIA superstring
on the resolved conifold where the D6-branes replaced by
$N$ units of R-R two-form fluxes through the two-sphere (${\bf S}^2\sim\cp^1$)
in the resolved conifold. This duality thus offers a way of understanding
the same physics at strong coupling. \\
{\bf \it M-theory interpretation}\\
The large $N$ duality in type IIA superstring theory has also
been lifted to M-theory on a $G_2$ manifold  \cite{AMV,AW} 
where it is known
to give the  so-called flop duality. This gives an M-theory interpretation
 of the  type IIA  duality transition between the
geometry involving branes and the one involving fluxes.
 Unlike the duality in string
theory, the phase transition here is smooth and does not
correspond to a topology changing geometric transition.
 To see this,   consider M-theory on a
seven-dimensional manifold $X^7$ defined by
\begin{equation}
  \label{X7Eqn}
  |z_1|^2+|z_2|^2-|z_3| ^2-|z_4|^2=\rho,
\end{equation}
where $z_i$ are four complex variables and $\rho$ is a real parameter.
This equation defines an $\R^4$ bundle over $\bf S^3$  having $G_2$ holonomy group. We now have two geometries
related by a flop~\cite{AMV}. These are given by:
\begin{eqnarray}
  X^7_\rho &=&\{ (z_i)\in C^4:|z_1|^2+|z_2|^2-|z_3| ^2-|z_4|^2=\rho\} \nn\\
  X^7_{-\rho}&=&\{ (z_i)\in C^4: |z_1|^2+|z_2|^2-|z_3| ^2-|z_4|^2=-\rho\}
\end{eqnarray}
or, equivalently
\begin{eqnarray}
  X^7_\rho &=& {\bf S^3}(z_1,z_2)\times C^2 (z_3,z_4)\cong {\bf S^3} \times R^4 \nn \\
  X^7_{-\rho} &=& C^2 (z_1,z_2)\times {\bf S^3} (z_3,z_4) \cong \R^4 \times {\bf S^3}.
\end{eqnarray}
In order to obtain, for instance, the small resolution with RR
two-form flux, we need to identify the M-theory circle with the
circle of the Hopf bundle $\bf S^1$ over $\bf S^2$, this being equivalent
to identifying the ``11th'' circle with one of the toric geometry
actions of the $\bf S^3$ that we have discussed above. On the other
hand, if we choose the M-theory circle in $\R^4$ we get the
deformed conifold geometry. Furthermore, this $U(1)$ toric  action
turns out to have fixed points, giving rise to singularities that
will correspond to D6-brane charges~\cite{LV}.
\subsection{ Duality in type IIB }
The mirror version in type IIB superstring theory of type IIA
duality states that the scenario with $N$ D5-branes wrapped around
the two-sphere in the resolved conifold, is equivalent (for large
$N$) to three-form fluxes through the ${\bf S}^3$ of the deformed
conifold. This has been generalized to other Calabi-Yau threefolds
where the blown-up geometries involve several $\cp^1$'s
\cite{CIV,CKV,CFIKV,AgMV,G,LL,AABDS}.\\
{\bf \it F-theory interpretation}\\
So far we have reviewed how a type IIA
duality (namely, the large $N$ equivalence between a system of
D6-branes wrapped on the finite $\bf S^3$ of $T^\ast {\bf S^3}$ and a type
IIA background with 2-form flux on the $\bf S^2$ of the resolved
conifold and no branes) arises from an M-theory flop~\cite{AMV}. A
natural question is about the ``mirror flop'', i.e. about an
alternative description of the mirror version of this type IIA
large $N$ duality. Mirror symmetry between type IIA and type IIB
compactifications maps the D6/${\bf S^3}$ system to a system of
D5-branes wrapped on the $\bf S^2$ of the resolved conifold and,
correspondingly, maps the type IIA configuration with fluxes only
(no branes) to a type IIB background with only three-form fluxes
on the blown-up ${\bf S^3}$ of the deformed conifold.
Naively, the corresponding ``mirror flop'' could be described by using
mirror symmetry in M-theory compactifications.
In this section we are going to see that what would be the
``mirror flop'' is actually mirror symmetry in the base of a
fourfold which is a (trivial) elliptic fibration over the
conifold. Such a fourfold naturally describes an  F-theory
compactification \cite{Va}.   To do so, we will proceed in two steps.  First we note that the above
discussed conifold geometries are related by the local mirror
symmetry transformation.  To see this, consider the mirror
geometry of the small resolution.   Using (8-20), this is defined by solving the
following constraint equations
\begin{eqnarray}
  a_1 y_1+a_2 y_2+a_3 y_3+a_4 y_4&=&0\\
  y_1 y_2&=&y_3y_4
\end{eqnarray}
A solution of the last equation is given by $y_1=1,\; y_2=x,
\;y_3=y,\; y_4=xy$, and the mirror geometry becomes
\begin{equation}
  a_1+a_2 x+a_3 y+a_4xy=0.
\end{equation}
At first sight, this geometry looks quite different
to~(\ref{DeformedConifoldEqn}). However, one can relate it to the
deformed conifold by  taking   the following limit in the complex
moduli space
\begin{equation}
  a_1=\mu,\quad a_2= a_3=0, \quad a_4=1.
\end{equation}
Now, we add the quadratic term. Note that this procedure has no
influence on the physical  moduli space. In this way, one can see
that mirror symmetry acts as follows
\begin{equation}
  \R^4\times {\bf S^2} \longleftrightarrow \R^3\times {\bf S^3}
\end{equation}
and then it could be   interpreted as a breaking/enhancement of an
$U(1)$ toric action, as discussed in the previous section. \\
The second  step is  to  use this feature and   our result in the context of the mirror duality
between M-theory on $G_2$ manifolds and F-theory on elliptic
Calabi-Yau fourfolds  \cite{B}.  Indeed,   let us  consider the conifold geometries described above
as a mirror pair, and use the limit $\R^4=\R^3\times {\bf S^1}$  corresponding
 to the moduli space of one monopole in
  M-theory compactification.  In such a limit,  M-theory on~(\ref{X7Eqn}) is dual to
F-theory on a Calabi-Yau fourfold which is the product of $T^2$ and
the small resolution of the conifold. The type IIB meaning of this
F-theory compactification has to be then the mirror system of the
D6/${\bf S^3}$ configuration, i.e. D5-branes wrapped on the ${\bf S^2}$ of the
resolved conifold geometry. Now, if we apply toric mirror symmetry in
the base of the fourfold on which we are compactifying F-theory, we
are led to an F-theory compactification on $T^2\times T^\ast {\bf S^3}$.
This provides the F-theory description of the other side of the type
IIB transition (with only RR three-form fluxes on the $\bf S^3$). 
This procedure immediately gives us another description of the
M-theory flop of~\cite{AMV} in terms of mirror symmetry: if, as
above, we look at~(\ref{X7Eqn}) as an ${\bf S^1}$ fiber over the
deformed conifold, we find that the flop can be alternatively
described by mirror symmetry in the base of that fibration.

\section{Geometric  transition  in two dimensions}
 In the previous section, we have  presented  the brane/flux duality in
four dimensions which arises from the transition in
the Calabi-Yau conifold.  A natural question is  whether a similar description
 exists for the transition on  the $Spin(7)$ manifold compactification.   In this section,
 we shall give such a description using the results of type II superstrings compactified on
 the conifold and  the toric realization  of the geometric
transition of the $Spin(7)$ manifolds. Following \cite{BR}, we shall
consider the large $N$ limit  two-dimensional gauge theories obtained by considering  type II superstrings 
  propagating on 
 $Spin(7)$ manifolds. The idea is to  discuss the
consequences of adding $N$ wrapped D-branes
to the set-up before letting the manifold undergo
the geometric transition. 
In the transition from the resolved to the deformed
$Spin(7)$ manifold, we initially have D-branes wrapping
$\cp^2$ (and its constituent two-spheres). We conjecture
that they are replaced, under the transition,
by R-R fluxes through ${\bf S}^5$ (and its constituent
three-spheres). Similarly in the transition from deformed
to resolved $Spin(7)$ manifolds, we conjecture that
D-branes wrapped around ${\bf S}^5$ (and its
constituent three-spheres) are replaced by R-R fluxes
through $\cp^2$ (and its constituent two-spheres).
The kind of D-branes involved and the more detailed
phase transition depend on which type II superstrings
are propagating on  $Spin(7)$ manifolds.
In what follows,  we shall therefore consider type IIA and type IIB
separately  leading  to different brane/flux
dualities.  First we consider type IIA geometry. Then we discuss the type IIB  modeñ  
 with  several D-brane configurations involving D1, 
D3, D5 and D7-branes.

\subsection{Duality in type IIA }
 First we  consider  type IIA superstring  on the deformed
$Spin(7)$ manifold. A two-dimensional
$U(N)$ gauge theory can be obtained by wrapping $N$
D6-branes around ${\bf S}^5$. The volume of
${\bf S}^5$ described by $r$ (\ref{r}) is proportional to the
inverse of the gauge coupling squared. This gauge model
has   only one supercharge. Thus   we have  ${\cal N}=1/2$.
At the transition point, the D6-branes disappear and are
replaced by R-R two-form fluxes through the two-spheres
embedded in $\cp^2$ in the resolved $Spin(7)$ manifold.\\
{\bf \it M-theory interpretation}\\
At this level, a natural question is about the analog of the flop
duality
 in four dimensions. The answer of this question  may  be given in terms  
 of  M-theory compactifications \cite{BR}.
 Indeed,  consider a nine-dimensional manifold
$X_9$ with an  $U(1)$ isometry. M-theory compactified on $X_9$ is
then equivalent to type IIA superstring  compactified on
$X_9/U(1)$. We start with the resolved $Spin(7)$ manifold  and identify the extra eleventh compact dimension of
M-theory with the ${\bf S}^1$ that generates (\ref{theta}). In
this way, the extra M-theory circle becomes the fiber in the
definition of ${\bf S}^5$ as a ${\bf S}^1$ fibration over
$\cp^2$. We thus end up with an $\R^4$ bundle over ${\bf S}^5$ as
the compactification space in M-theory. As a consequence, the
moduli space of M-theory on such a background is parameterized by
the the real parameter $r$ defining the volume of ${\bf S}^5$
(\ref{r}), and cannot be complexified by the C-field. Starting
with the resolved $Spin(7)$ manifold, on the other hand, the
eleventh M-theory dimension is obtained by extending $\R^3$ to
$\R^4$ with the isometry being a trivial $U(1)$ action on the
fiber $\R^4$. Using arguments similar to those in \cite{AMV}, it is
conjectured \cite{BR} that this lift to M-theory gives rise to a (smooth)
flop transition in the $\R^4$ bundle over ${\bf S}^5$ where a
five-sphere collapses and is replaced by a new five-sphere. In our
scenario, however, the physics resulting from the type IIA
superstring compactification undergoes a singular phase transition
due to the absence of 5-form gauge field in the spectrum.

\subsection{Duality in type IIB }
 Now   we consider type IIB superstring on the resolved
$Spin(7)$ manifold. 
In  this case,  two dimensional  gauge models
 can be  engineered  using  several  brane configurations. In particular,
we have the following D-brane realizations:
\begin{itemize}
\item{ $N$ parallel  D1-branes filing the two-dimensional spacetime ( and placed at the singular
point of the $Spin(7)$  manifold).}  This gives  an $ U(N)$ gauge symmetry in spacetime.
\item{ $N$ parallel wrapped D3-branes  over  real  2-cycles   and  filing the 
two-dimensional spacetime ( and placed at the singular
point of the $Spin(7)$  manifold). This leads aslo to  $2D$ $ U(N)$ theory.}
\item{ One may  also have systems involving D3-branes and  D1-branes. This may lead to $  U(N_1)\times U(N_2)$ product 
gauge symmetry.}
\item{ $N $ parallel  wrapped D5-branes  over  $\cp^2$  and  filing the 
two-dimensional spacetime  ( and placed at the singular
point). This leads to $ U(N)$ gauge symmetry.}
\item{ $N$  parallel wrapped  D7-branes   over    real  6-cycles  in  $Spin(7)$  manifold and  filing the 
two-dimensional spacetime ( and placed at the singular
point). This gives  an $ U(N)$ gauge symmetry.
} 
\item{ One may  also have systems involving  D7-branes, D5-branes,  D3-branes and D1-branes. 
This may lead to $ U(N_1)\times U(N_2) \times U(N_3) \times U(N_4)$ product 
gauge symmetry.}
\end{itemize}
 To  realize  two last configurations,  we  need a 6-cycle. To get  that, we should mod out, the resolved  
$Spin(7)$  manifold, by a $Z_2$  discrete group acting on the fiber direction  of $\R^4$ bundle over $\cp^2$.  This corresponds to ALE space with 
$A_1$ singularity over $\cp^2$. After the deformation of this singularity, one gets a 6-cycle
 given  by $\cp^1\times \cp^2$. In  shrinking limit, this will correspond to 6-cycle/5-cycle transition 
in superstring compactifications. This  looks like the case  of 4-cycle/3-cycle transition studied in \cite{CFIKV}.
In particular, after the transition,  one may have  ${\bf S}^5/Z_2$
as a dual cycle.\par
Now return to the  large $N$ limit duality in type IIB.  Here, we are not going to 
deal with  the all  above  brane configurations.   However, we 
consider  a system with  
only D5-branes. Then we discuss a  case where we have  D3-branes.  Since the type IIB  superstring does
not support four-forms, one can wrap  D5-branes around $\cp^2$. As we have seen,
a two-dimensional gauge model with  $U(N)$ gauge symmetry can be obtained
by wrapping $N$ D5-branes on $\cp^2$. The volume of
$\cp^2$ described by $r$ (\ref{r}) is proportional to the
inverse of the gauge coupling squared. This two-dimensional
model has only one supercharge so ${\cal N}=1/2$.
Now, when the manifold undergoes the geometric transition
to the deformed $Spin(7)$ manifold, the $N$ D5-branes
disappear and we expect a dual physics with $N$ units of
R-R three-fluxes through the compact three-cycles, ${\bf S}^3$,
in the intersecting Calabi-Yau threefolds.
These fluxes could be accompanied  by  some NS-NS fluxes
through the non-compact dual three-cycles in the six-dimensional
deformed conifolds. In order to handle the associated divergent
integrals, one would have to introduce a cut-off to regulate
the infinity \cite{CIV}. 
\par
 So far, we have studied the  geometric transition  involving only  D5-branes  on the  $Spin(7)$ manifold.
Now we  would like  to  go beyond this model by  adding, for instance, D3-branes. We will show
 that this  brane configuration could 
  be related to del Pezzo surfaces. 
Here,  our  construction of  such   brane gauge models will  be based on 
 ${\cal N}=2$ sigma model  realization of  the internal  compact geometry.  
 For simplicity, let us consider $dP_1$. Indeed, this can be realized  by   $U_1(1) \times U_2(1)$ gauge theory  
  with  four chiral fields $\phi_i$ with charges 
\begin{eqnarray*}
Q_{(1)}=(1,1,1,0)\\
Q_{(2) }=(0,0,1,1)
\end{eqnarray*}
 The vanishing condition of  D-terms read as
\begin{eqnarray*}
  |\phi_1|^2+|\phi_2|^2+|\phi_3| ^2&=&\alpha_1 \\
   |\phi_3| ^2+|\phi_4|^2&=&\alpha_2
\end{eqnarray*}
 and define   $dP_1$.   In what follows we will identify $U_1(1) \times U_2(1)$  sigma model gauge
 group with the gauge symmetry living the worldvolume of D-branes.
Indeed,  omitting  first $U_2(1)$ and after the  wrapping procedure,    $U_1(1)$ gauge symmetry can be identified with
 an $U(1)$  gauge field  living on the D5-brane worldvolume \footnote {This  model   can be identified 
 with the one studied in  the   previous  subsection.}. However,  
 due to the presence  of $U_1(2)$,  we really  could  introduce  an  extra brane. In this
way,  $U_2(1)$   can  be identified  with the $U(1)$ gauge field  living on its worldvolume.
Since $U_2(1)$,   in ${\cal N}=2$ sigma model,  corresponds  to a  blowing-up   2-cycle,  we should add a D3-brane. 
The latter  can wrap  such a  cycle and leads to   an extra $U(1)$  factor in two-dimensional spacetime.
  So, the  general gauge symmetry  is then $U(1) \times U(1)$. 
In what follows,   we  should  note  the following points:\\
1. In the vanishing limit of the 2-cycle,  where $dP_1$ reduces to   $\cp^2$, one gets only one U(1) factor  
 corresponding to one D5-brane.\\
2. The above  model  could be generalized for the case where we have  more than one
  brane. If we assume that one  has the
 same number of D5 and D3-branes, then 
 the gauge group should be $U(N) \times U(N)$.\\
3. After the transition point, these branes disappear and replaced by 3-form and 5-form fluxes on
  on ${\bf S}^5/Z_2$.
\par
We conclude this work  with a comment regarding a  F-theory interpretation of type IIB
geometric transition  in the compactification on  the above $Spin(7)$ manifold.
This  is given by   a F-theory compactification  on a ten-dimensional
manifold  involving a ${\bf S}^5$  flop transition. While we were thinking on this question after
   completing the first version of this work,  \cite{TT}  appeared which has a 
considerable overlap with this comment. In \cite{TT},   a
11-dimensional interpretation  has been  given. However, here we give a
direct compactification on a ten-dimensional manifold preserving
only one supercharge in two dimensions.
Based on  a  close analogy to the Calabi-Yau case, we propose the following
ten-dimensional
\begin{equation}
T^*({\bf S}^5)/\sigma,
\end{equation}
where $T^*({\bf S}^5)$ is the complex deformation of  the singular Calabi-Yau
five-folds given by
\begin{equation}
w_1^2+w_2^2+w_3^2+w_4^2+w_5^2+w_6^2=0.
\end{equation}
This manifold has  nice  features   supporting our proposition:  \\
(1) In  string theory compactifications, $T^*({\bf S}^5)$ preserves $1/16$  of initial supercharges
and in the presence of $\sigma$  it should be
$1/32$. Thus, F-theory on the above ten-dimensional manifold leads to only
one supercharge in two dimensions.\\
(2) The presence of $\sigma$  may lead to a   ${\bf S}^5$  flop
transition in the quotient space. It  is easy  to see this by
taking  $w_j$ as $x_j+iy_j$ and rewritting the above algebraic equation. In this way,   $T^*({\bf S}^5)$  can be
described by
\begin{equation}
{\bf x}\cdot{\bf x }- {\bf y}\cdot{\bf y}=r,\ \ \ \
  {\bf x}\cdot{\bf y}=0
\label{base}
\end{equation}
where $r$ is a real parameter describing the size of ${\bf S}^5$.  If we think
   that $\sigma$ acts as $x \to y$, and  $ r\to -r$, then we have two  spheres being
     connected by the so-called a flop transition. \\
(3) After  a compactification, the  reduced manifold may involve three Calab-Yau conifolds.
To see this let us  take a
simple linear coordinates from $(w_1, \ldots,w_6)$ to  $(z_1,z_2,z_3,z_4,u,v)$. This allows
one to describe the above manifold by
\begin{equation}
\sum_{1\leq i<j\leq 4}z_iz_j=z_4( z_3+z_2+z_1)+z_3(z_2+z_1)+z_2z_1=uv
\label{trans}
\end{equation}
such that
\begin{equation}
{\left(\begin{array}{ll}w_1\\ w_2\\w_{3} \\w_{4}\\ w_{5}\\  w_{6}
  \end{array}\right)} =\ \left(\begin{array}{lllllll} \sqrt{3}&i&i&i&0&0\\
  i&\sqrt{3}&i&i&0&0\\
   i&i&\sqrt{3}&i&0&0\\ i&i& i&\sqrt{3}&0&0\\
    0&0&0&0&d&-id\\
    0&0&0&0&-id&d   \end{array}\right){\left(\begin{array}{ll}
     z_1\\  z_{2}\\z_{3}\\z_{4} \\ u\\ v
  \end{array}\right)}
\end{equation}
with
\begin{equation}
d\ =\ \sqrt{\frac{2+\sqrt{3}}{2}}
   \ +\ i  \sqrt{\frac{2-\sqrt{3}}{2}}.
\label{d}
\end{equation}
Equation (\ref{trans}) describes six Calabi-Yau conifolds given by  $uv=z_iz_j;
z_k=z_\ell=0, i\neq j\neq k\neq\ell$. However if we identify one
of the $z_i$  complex variables with the   $T^2$  torus of
F-theory compactification, this number reduces to three and it is
in agrement with the $spin(7)$ geometry involving three Calabi-Yau
manifolds.\\
 (4)  Finally,  the topology of $spin(7)$  manifold can be
obtained  by identifying the $T^2$  torus of F-theory with the
toric actions of  $T^*({\bf S}^5)/\sigma $. Using the results of section
2 and \cite{AB},  one can get the resolved  $spin(7)$ by the choosing
one circle in the $S^5$ base geometry and one circle in the
corresponding cotangent directions, while the deformed $spin(7)$ manifold is
obtained by  identifying the two F-theory circles with toric
actions of
the cotangent bundles.

\section{ Discussions}

In this study,  geometric transitions in type II superstrings
 on  Calabi-Yau conifold  and
 $Spin(7)$ manifolds  have  been discussed  using
   toric geometry. In the Calabi- Yau case, we have  proposed a new 
 F-theory interpretation  for type IIB propagating on the conifold.
 Following \cite{BR} on $Spin(7)$ manifolds,   we have given a picture  
 for understanding the 
 topology changing transition in the $Spin(7)$ manifolds in terms of  the
Calabi-Yau conifold transition. Then,  we have studied   
brane/flux dualities  in two dimensions using several  brane configurations.  
This  study   give  first
examples of geometric transitions with  ${\cal N}=1/2$
supersymmetry. Then, an interepretation in terms of F-theory has been given. \par

This work opens up for further  discussions.  We shall collect some of them.
\begin{itemize}
\item{ In this study, we have considered  the $Spin(7)$ manifold as intersecting Calabi-Yau
threefolds over a triangle where the $Spin(7)$ transition
corresponding  to three simultaneous conifold transitions.  It would be interesting to study
 the geometries associated to individual conifold
transitions.} 
\item{ It should be nice to find a sigma model explanation of the $Spin(7)$ transition.} 
\item{ We expect that all studies  which have been  done for the Calabi-Yau
conifold could be pushed to   $Spin(7)$ manifolds. It would be nice to study the analogue of  the
Klebanov-Witten model on the conifold \cite{KW} for $Spin(7)$ manifold. }
\end{itemize}
 We leave these questions for future works.
\\[.3cm]
{\bf Acknowledgments.}   I would like to thank   Cesar Gomez,  Karl Landsteiner, Ernesto Lozano
 Tellechea, J{\o}rgen Rasmussen and   El Hassan Saidi, for
discussions, collaborations and  scientific  helps.   I am very grateful for the support I have received 
from CSIC.  This work  is supported by
Ministerio de Educaci\'{o}n Cultura y Deportes (Spain) grant SB 2002-0036.

\end{document}